# The Structure and Dynamics of a Bright Point as seen with Hinode, SoHO and TRACE

D. Pérez-Suárez[1], R.C. Maclean[1,3], J.G. Doyle[1], and M.S. Madjarska[2]

[1] Armagh Observatory, College Hill, Armagh BT61 9DG, N. Ireland
[2] Max-Planck-Institut für Sonnensystemforschung, Max-Planck-Str. 2, 37191 Katlenburg-Lindau, Germany
[3] now at: School of Mathematics and Statistics, University of St. Andrews, North Haugh, St. Andrews, Fife KY16 9SS, UK

**ABSTRACT**

*Context.* Solar coronal bright points have been studied for more than three decades, but some fundamental questions about their formation and evolution still remain unanswered.
*Aims.* Our aim is to determine the plasma properties of a coronal bright point and compare its magnetic topology extrapolated from magnetogram data with its appearance in X-ray images.
*Methods.* We analyse spectroscopic data obtained with EIS/Hinode, Ca II H and G-band images from SOT/Hinode, UV images from TRACE, X-ray images from XRT/Hinode and high-resolution/high-cadence magnetogram data from MDI/SoHO.
*Results.* The BP comprises several coronal loops as seen in the X-ray images, while the chromospheric structure consists of tens of small bright points as seen in Ca II H. An excellent correlation exists between the Ca II H bright points and increases in the magnetic field strength, implying that the Ca II H passband is a good indicator for the concentration of magnetic flux. Furthermore, some of the Ca II H bright points are the locations of the loop foot-points as determined from a comparison between the extrapolated magnetic field configuration and the X-ray images. Doppler velocities between 6 and 15 km s$^{-1}$ are derived from the Fe XII and Fe XIII lines for the bright point region, while for Fe XIV and Si VII they are in the range from $-15$ to $+15$ km s$^{-1}$. The coronal electron density is $3.7\ 10^9$ cm$^{-3}$. An excellent correlation is found between the positive magnetic flux and the X-ray light-curves.
*Conclusions.* The remarkable agreement between the extrapolated magnetic field configuration and some of the loops composing the bright point as seen in the X-ray images suggests that a large fraction of the magnetic field in the bright point is close to potential. However, some loops in the X-ray images do not have a counterpart in the extrapolated magnetic field configuration implying a non-potential component. The close correlation between the positive magnetic flux and the X-ray emission suggests that energy released by magnetic reconnection is stimulated by flux emergence or cancellation.

**Key words.** Sun: activity – Sun: magnetic fields – Sun: corona – Sun: chromosphere – Line: profiles

## 1. Introduction

Coronal X-ray bright points (BPs) were first observed in rocket images in 1969 (Vaiana et al. 1973) and were seen as diffuse clouds with a bright core, although when viewed at higher spatial resolution, small loops are resolvable (Sheeley & Golub 1979). BPs are coronal structures smaller than 60″ that are associated with the interaction of photospheric bipolar magnetic features. Up to two third of them are related to cancellation of pre-existing magnetic features rather than to the emergence of new magnetic flux (Webb et al. 1993). McIntosh & Gurman (2005) recently reported on bright point statistics showing that 100 times more BPs were observable in EIT 171 Å than in the 284 Å filter. They speculated that there is a temperature dependence in the generation mechanism. The lifetime of an individual BP can be up to ≈40 hours (Golub et al. 1974).

Habbal & Withbroe (1981) used Skylab data to show that BPs exhibit large variations in their emission in chromospheric, transition region and coronal lines. This work was followed up more recently by Madjarska et al. (2003), Ugarte-Urra et al. (2004) and Ugarte-Urra et al. (2005) who studied BPs at transition region temperatures and derived their plasma characteristics such as electron density variability, Doppler shifts and intensity oscillations. The magnetic structure of BPs has been modelled by various authors (Parnell et al. 1994; Longcope et al. 2001; Brown et al. 2001; von Rekowski et al. 2006a). The general view is that these features result from magnetic reconnection, although identifying the specific type of reconnection involved is still a challenge. Coronal reconnection begins when opposite polarity magnetic fragments approach one another, with the resulting release of energy into the corona leading to the formation of a BP.

With the launch of Hinode (Kosugi et al. 2007), new frontiers have opened for studying coronal BPs in combination with data from the Solar & Heliospheric Observatory (SoHO; Domingo et al. 1995) and the Transition Region and Coronal Explorer (TRACE; Handy et al. 1999). Here we present the results of a multi-spacecraft/multi-instrument study of a BP. We use the EUV Imaging Spectrometer (EIS/Hinode; Culhane et al. 2007) together with Ca II H and G-band images from the Solar Optical Telescope (SOT/Hinode; Tsuneta et al. 2007), EUV images from TRACE and the Extreme ultraviolet Imaging Telescope (EIT/SoHO; Delaboudinière et al. 1995), and X-ray images from the X-ray Telescope (XRT/Hinode; Golub et al. 2007; Kano et al. 2007) as well as high-resolution/high-cadence magnetograms from the Michelson Doppler Imager (MDI/SoHO; Scherrer et al. 1995).

With the high-cadence magnetic field data coupled with images from XRT plus Ca II H and G-band data, we look at the structure of the bright point at photospheric, chromospheric and coronal heights to determine the location of sites of energy dissipation (§ 3). This is then combined with EIS spectral data on transition region and coronal lines to determine the flow, temperature and electron density structure of the bright point. Finally, in § 4, a magnetic charge topology (MCT) analysis (Longcope 2005) is used to study the structure and evolution of the three-dimensional (3D) magnetic field associated with the bright point. These extrapolations are then compared with the appearance of the BP in XRT images, and the location of the Ca II BPs is compared with that of the foot-points of the extrapolated magnetic field lines.

**Table 1:** Description of the analysed data.

| Instr | Filter | Pixel sz.* | Exp. (s) | Cad. (s) | FOV |
|---|---|---|---|---|---|
| EIT | 195 Å | 5.26″ | 5 | 720 | Full Disk |
| MDI | Ni I 6767 Å | 0.6″ | 30 | 60 | 620″ × 305″ |
| TRACE | 1550 Å | 0.5″ | 9.74 | 25 | 388″ × 388″ |
| SOT | Ca II H<br>G-Band | 0.11″ | 0.15<br>0.05 | 30 | 113″ × 56″ |
| XRT | Al-Mesh | 1.03″ | 16.4 | 20 | 395″ × 395″ |
| EIS | 9 spec win | 2.02″×1″ | 45 | - | 75″ × 512″ |

*1″ corresponds to ≈715 km on the Sun from SoHO view.

**Table 2:** A list of the various spectral lines present in the nine EIS/Hinode spectral windows for the rasters.

| | Window | $\Delta\lambda$(Å) | Lines Identified |
|---|---|---|---|
| 1 | Fe X 184.54 | 2.29 | O VI 184.12<br>Fe X 184.54<br>Fe VIII 185.21 |
| 2 | Fe XII 186.88 | 1.23 | Fe VIII 186.60<br>Fe XII 186.88 |
| 3 | Ca XVII 192.82 | 2.83 | Fe XII 192.39<br>Fe XI 192.83<br>Fe XII 193.51 |
| 4 | Fe XII 195.12 | 1.23 | Fe XII 195.12 |
| 5 | Fe XIII 202.04 | 1.22 | Fe XIII 202.04 |
| 6 | Fe XIII 203.83 | 1.23 | Fe XII 203.72<br>Fe XIII 203.83 |
| 7 | He II 256.32 | 2.65 | He II 256.32<br>Fe XII 256.94<br>Fe X 257.26 |
| 8 | Mg VI 270.40 | 1.22 | Mg VI 270.39<br>Fe XIV 270.52 |
| 9 | Si VII 275.35 | 1.22 | Si VII 275.35<br>Si VII 275.67 |

## 2. Observations and data calibration

The data analysed here were obtained on 2007 April 13 between 16:30 UT and 23:54 UT. We used two types of instrument: imagers and spectrographs. The characteristics of the data obtained from the imagers are outlined in Table 1.

Three rasters and two time series were obtained from EIS/Hinode with the study "BP_SUMER_EIS". The rasters were made by stepping the 2″ slit in 1.87″ increments across an area of 75″ × 512″ around the centre of the solar disc with an exposure time of 45 s. Detailed information about the observed spectral lines can be found in Table 2. The BP studied here was not in the field-of-view of the time series. Hence, we only use the raster data. All the data have been reduced and calibrated with the standard procedures as given in the SolarSoft (SSW)[1] library. Further details may be obtained from the instrument webpages.

### 2.1. Alignment and data correction

All images from TRACE and Hinode were converted to SoHO view (L1) and co-aligned as described below. The first step was to remove any jitter in each instrument by cross-correlating all the images with the mean image. Then all data were aligned with respect to the TRACE 1550 Å images. First we cross-correlated TRACE 1550 Å with the SOT Ca II H images and found a variable offset from 4″ to 10″ in the E–W direction (Solar X), and from 10″ to 15″ in the N–S direction (Solar Y). We then aligned the EIS He II 304 Å raster images with SOT Ca II H and TRACE 1550 Å. The next step was to cross-correlate EIS Fe XII 195 Å images with XRT, and XRT with EIT. We obtained offsets between SOT and XRT close to those derived by Shimizu et al. (2007) i.e. 2″ in Solar X and 14″ in Solar Y. Between EIS and SOT we obtained values similar to those shown in *EIS-wiki*[2], i.e. 17″ in Solar X and 20″ in Solar Y with the long-wave CCD. The internal offsets between the two CCDs within EIS are known (Young et al. 2007b). MDI was then cross-correlated with Ca II H and TRACE 1550 Å. All images were rotated to a common time chosen to be 19:00 UT.

The temperature response of the XRT Al-mesh filter has a double peak, with the first peak below $10^6$ K and the second one at $\approx 5 \times 10^6$ K. The peak response of the TRACE 171 Å filter is around $10^6$ K, thus we compared our five TRACE 171 Å images with the XRT images in order to examine the extent of the overlap and therefore gain some insight into the temperature of the BP. The XRT images show a central brightening, whereas in TRACE 171 Å the bright area is at least $\approx 50\%$ smaller, thus indicating that the bulk of the BP has a temperature structure above $10^6$ K.

The EIS data have two major instrumental effects which have to be corrected: a tilt in the slit on the CCD and a sinusoidal drift of the lines on the detector due to orbital changes. The orbital variation is of the order of 1.3 spectral pixels ($\approx 45$ km s$^{-1}$), while the tilt correction amounted to $\approx 77$ km s$^{-1}$ over the length of the slit for the strong Fe XII 195.12 Å line.

## 3. Results and discussion

### 3.1. The Bright Point's Global Appearance

The bright point was first identified in the EIT and XRT images, which showed a structure of 20″ × 40″. The BP first appeared on

---
[1] http://www.lmsal.com/solarsoft/
[2] http://msslxr.mssl.ucl.ac.uk:8080/eiswiki/Wiki.jsp

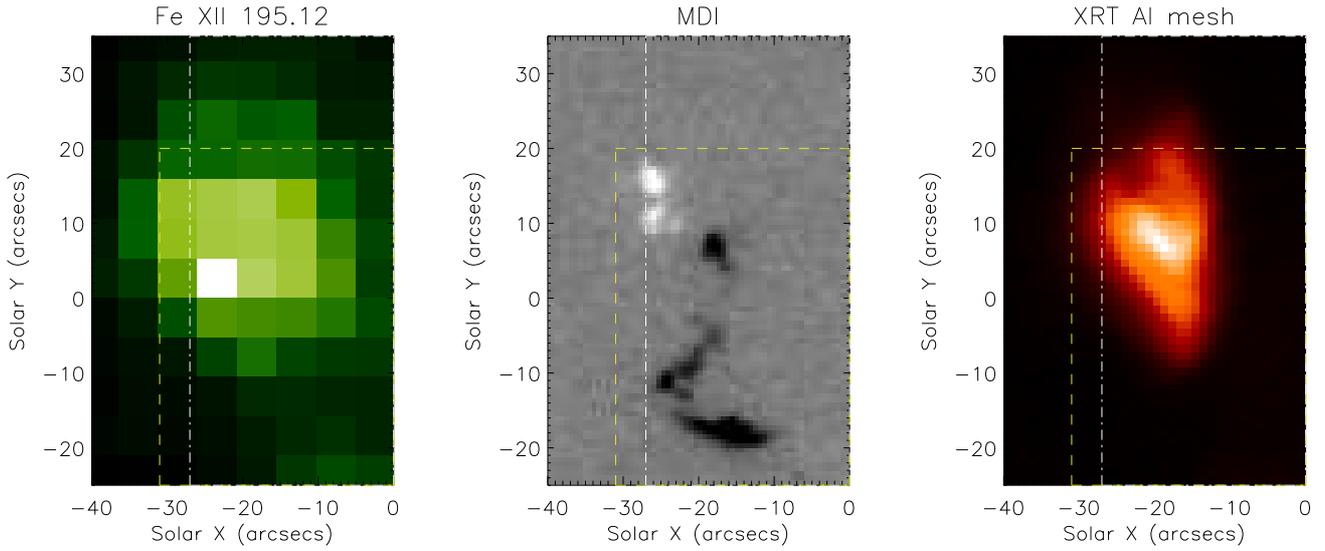

**Fig. 1:** Images of the bright point as seen on 13 April 2007. Left to right: EIT (22:00:11 UT), MDI (22:01:01 UT) and XRT (22:01:05 UT). The outlined dashed box is the FOV as shown in Fig. 2, while the dot-dashed box shown in Fig. 3. See movie1 online.

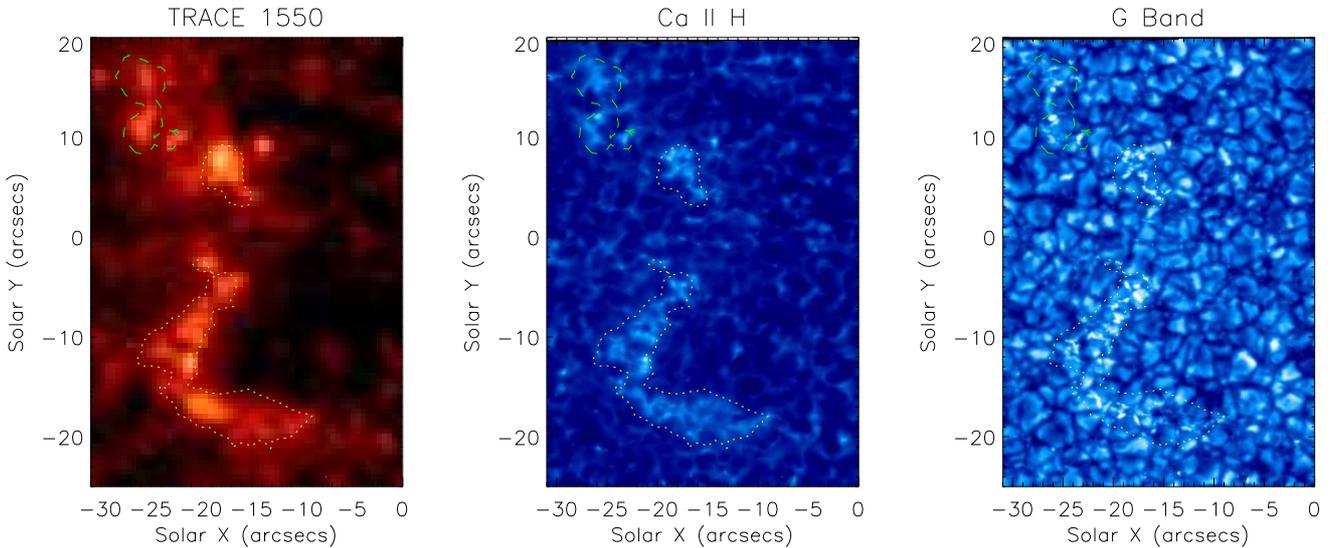

**Fig. 2:** Images of the bright point (overlaid with MDI magnetic field contours of ∼ ±150 G) as seen on 13 April 2007. From left to right: TRACE 1550 Å filter (22:00:12 UT), SOT Ca ɪɪ H (22:01:10 UT) and SOT G-band (22:01:06 UT) , see movie2 online.

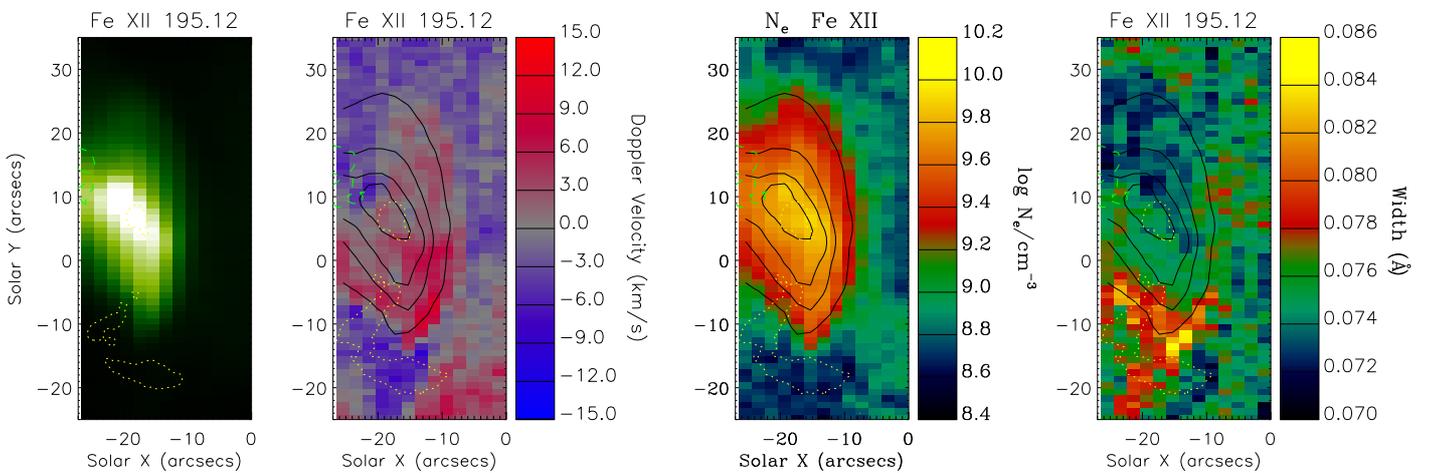

**Fig. 3:** Images of the bright point as seen on 13 April 2007 showing the intensity in Fe xɪɪ 195.12 Å, the Doppler velocity as measured by Fe xɪɪ 195.12 Å, the electron density as measured by the Fe xɪɪ (186.88+186.85)/(195.12+195.18) line ratio and the FWHM obtained in Fe xɪɪ 195.12 Å. The solid lines are the intensity contours, while the dashed and dotted lines are the MDI magnetic field contours ∼ ±150 G. These data were taken between 22:01:14 UT and 22:32:31 UT.

12 April 2007 (EIT 195 Å) at ≈22:24 UT and disappeared on 14 April 2007 at ≈13:13 UT resulting in a lifetime of approximately 37 hours. The general shape of the BP as viewed with EIT was almost constant throughout its lifetime (Figure 1, left panel). In XRT, the overall structure of the BP did not change significantly throughout its lifetime, although small changes occurred which will be discussed later in the paper.

Another factor confirming that this feature is a bright point is its magnetic field configuration. EUV/SXR bright points are generally associated with two interacting magnetic polarities of opposite sign. As we can see in Fig. 1 (middle panel), this BP is dominated by a negative field, and as Parnell et al. (1994) predicted, it looks brighter over the area with the stronger magnetic flux. The BP was created following the emergence of a small positive magnetic patch on 12 April at ≈20:48 UT into a region dominated by negative field patches. Unfortunately, we do not have high-resolution MDI data until 20:56 UT the next day, so we are unable to comment on the magnetic field evolution during this period of time. In the XRT images (Figure 1, right panel), the BP is seen as a blurred feature, however, when image sharpening is applied, it is possible to distinguish individual loop structures that evolve in time (see later). These loops connect the opposite polarities seen in MDI. We also observe a jet-like feature that appears between the two closest bipolar structures, with a flux intensity 3 to 30 times stronger than the quiet Sun value. A movie sequence of the EIT, MDI and XRT data can be viewed at movie1

The high-resolution MDI images give a detailed view of the BP's magnetic field configuration. Its shape is that of an inverted question mark throughout the period of the MDI observations (2 hours), but in detail there are some small magnetic patches that appear, disappear, merge and split. This will be discussed in more detail in § 4.

### 3.2. The Bright Point in TRACE 1550 Å, Ca II H and G-band

The BP was observed in the lower atmospheric layers with TRACE 1550 Å and SOT/Hinode Ca II H and G-band (Figure 2 left, middle and right panel respectively). All of them show a general shape consistent with the location of the strong magnetic field regions.

As seen by TRACE, the BP consists of around 50 small bright point-like features with a size of $1'' \times 1''$. Several are grouped together forming a large feature while others are isolated. These latter ones are less quiet. Since the TRACE 1550 Å bandpass is rather wide it is difficult to say whether the detected variability happen at chromospheric or transition region temperatures, although previous studies suggest that in features such as BPs, the C IV 1548/1550 Å lines should dominate.

G-band images show a good correlation with what is seen in TRACE 1550 ÅThe brightenings appear at the edges of the granules, moving around them. Finally, in Ca II H, the brightenings are just one third brighter than the background. A detailed observation of these small chromospheric BPs in space and time shows, when they are compared with high-resolution MDI magnetograms, that they are a good indicator for concentration of the magnetic flux (see movie3).

### 3.3. Spectroscopic analysis

EIS observed the BP in three rasters starting at 18:25 UT, 20:13 UT and 22:01 UT. Each raster lasted ≈ 31 mins. We fitted the Fe XII 186.88 Å and Fe XIII 202.04 Å lines with a single

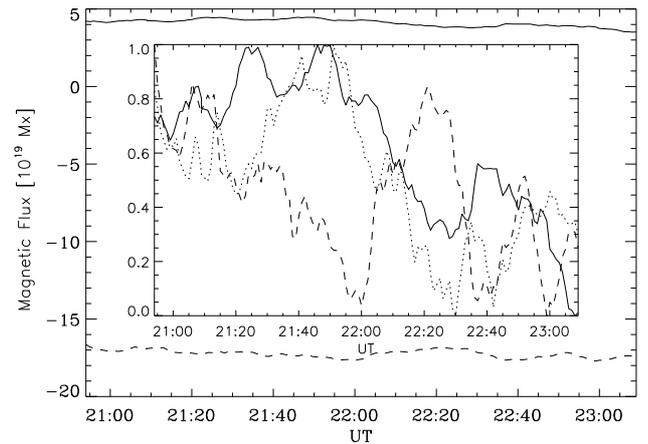

**Fig. 4:** Time variation of the magnetic fluxes of the two main BP magnetic polarities. The solid line indicates the positive polarity, and the dashed line indicates the negative polarity. In the inner panel both fluxes are normalized. The dotted line shows the XRT flux of the BP.

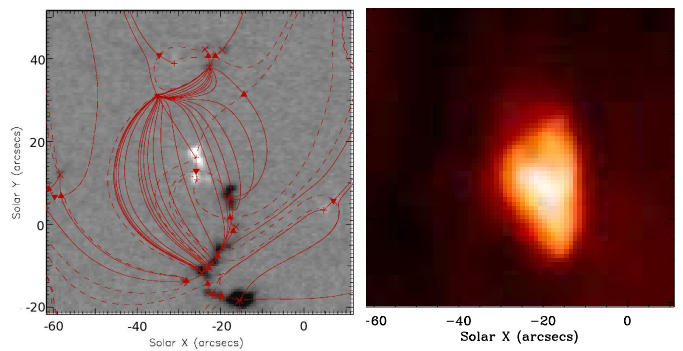

**Fig. 5:** A plan view of the topology of the BP region at 21:14:02 UT on 13 April 2007 overlaid on the MDI image. "+" represents positive magnetic sources, "×" represents negative sources, upwards-pointing triangles represent positive magnetic null points, and downwards-pointing triangles represent negative null points. In the photospheric plane, solid fieldlines are spines, and dashed fieldlines are the intersection of separatrix surfaces with the photosphere. 3D solid magnetic field-lines have also been plotted to show the important separatrix surface in the topology: the dome of field-lines contains all the magnetic flux in the BP. The brightening in XRT shows a similar outline.

Gaussian; for Fe XII 195.12 Å and Fe XIII 203.80 Å double and triple Gaussians were used, as justified by Young et al. (2008). For every pixel we obtain the total intensity, the line centre and the line width. Figure 3 (1$^{st}$ panel) shows the 21:01 UT raster as seen in Fe XII 195 Å, while Figure 3 (2$^{nd}$ panel) shows the corresponding Doppler velocities. The electron density as determined with the Fe XII 186/195 line ratio (after fixing the grating tilt (Young et al. 2008)) is shown in Figure 3 (3$^{rd}$ panel) and the line width of Fe XII 195Å is given in Figure 3 (4$^{th}$ panel). The Doppler velocities were calculated relative to the whole dataset. It is clear that the South-East section of the bright point has a predominant redshift.

In Table 3 we show the Doppler velocities of the BP obtained from a single spectrum averaged over the whole BP. The general trend is for the lines to be red-shifted. If we look in detail, however, the range of velocities in the BP varies between differ-

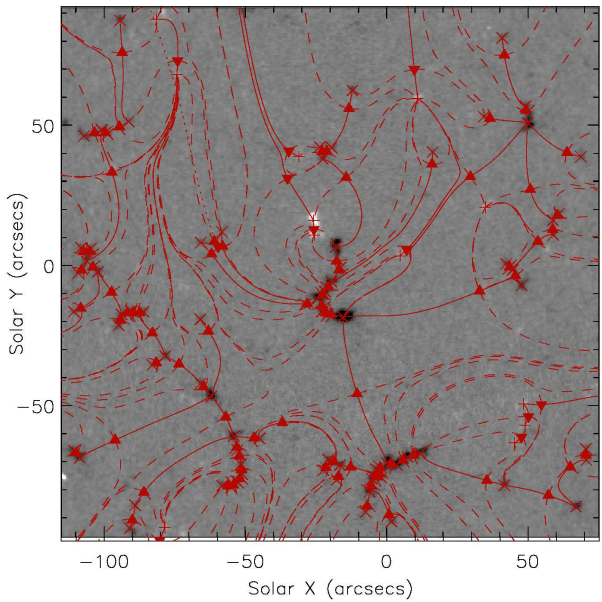

**Fig. 6:** The photospheric footprint of the magnetic topology of the BP region at 21:14:02 UT (overlaid on the MDI image), showing sources and null points as in Figure 5. The characteristic shape of the BP is apparent, centred at $(-20, 0)$. Spine field-lines are also shown as solid curves, and the intersections of the separatrix surfaces with the photosphere are shown as dashed curves.

**Table 3:** Mean Doppler shift of the BP as a whole along the three rasters.

| Line | | Log T | Doppler shift [km s$^{-1}$] |
|---|---|---|---|
| Fe x | 184.54 | 6.0 | 0.2 |
| Fe xii | 195.12 | 6.2 | 1.2 |
| Fe xiii | 202.04 | 6.2 | 2.3 |
| He ii | 256.32 | 4.7 | 6.4* |
| Fe xiv | 270.52 | 6.3 | 1.8 |
| Si vii | 275.35 | 5.8 | 0.2 |

* He ii is blended $\sim 20\%$ with three other lines in the red side.

ent locations. We obtain for Fe x 184.54 Å values in the range from $-10$ to $+10$ km s$^{-1}$, for Fe xii 195.119 Å values from $-8$ to $+8$ km s$^{-1}$, whereas for the Fe xiii 202.04 Å there is a red shift in the most of the pixels within a range of 6 to 18 km s$^{-1}$. In the case of He ii 256.32 Å, the values oscillate between $+5$ to $+25$ km s$^{-1}$ but it is not reliable because the line is blended on the red side by Si x 256.37 Å, Fe xiii 256.42 Å and Fe xii 256.41 Å (Young et al. 2007a). For the hottest line studied, Fe xiv 240.52 Å, we obtained values in a range $-15$ to $+15$ km s$^{-1}$. For the coolest line, Si vii 275.35 Å, we obtain a similar range.

Brosius et al. (2007) recently showed some Doppler shift from a BP. The values that we obtained are in the range which they showed for the same ions, i.e. He ii and Fe xiv. For Fe x 184.54 Å and Si vii 275.35 Å we obtain errors of $\sim 4$ km s$^{-1}$; for Fe xiv 270.52 Å the errors are $\sim 2$ km s$^{-1}$. For the rest of the stronger lines, the error is smaller, being $\sim 1$ km s$^{-1}$.

In addition to Fe xii 186.88+186.854/195.119 (note that S xi 186.83 Å contributes about 5% blending, Peter Young, private communication), we also used Fe xiii 203.797+203.828/202.04 with the atomic data for both ions from the CHIANTI database (Dere et al. 1997; Landi et al. 2006) to derive the electron density. This line pair is almost temperature insensitive; changing the formation temperature by $\pm 20\%$ produces less than a $\pm 10\%$ change in the electron density. For each raster, we selected an area of 15″ × 40″ over the BP (see Table 1). We used a quiet Sun area to determine the background emission. The results for one of the rasters are shown in Figure 3 (3$^{rd}$ panel). It is easy to see the structure of the bright point. Furthermore, there is an increment in the electron density in the core of the bright point. Excellent agreement is obtained between the two ratios, implying an electron density of $\approx 3.7 \times 10^9$ cm$^{-3}$ at coronal temperatures.

**Table 4:** Mean electron densities over the whole bright point derived from the Fe xii and Fe xiii line ratios for the three EIS rasters.

| Ion | Ratio | log $N_e$ (cm$^{-3}$) |
|---|---|---|
| Fe xii | $\frac{186.88+186.854}{195.119}$ | 9.55 |
| | | 9.55 |
| | | 9.58 |
| Fe xiii | $\frac{203.797+203.828}{202.04}$ | 9.52 |
| | | 9.56 |
| | | 9.49 |

The width of the Fe xii 195.12 Å line (Figure 3, 4$^{th}$ panel) shows an increment of 10 mÅ in the bottom part of the BP compared with the top and background. This is the location of the strong negative magnetic field. This is also the location of the largest line-shift, which may suggest a high degree of turbulence and/or additional unresolved flows in the loops.

## 4. Magnetic Topology

MDI took high-resolution line-of-sight photospheric magnetogram data of the bright point between 20:54 UT and 23:09 UT. This dataset of just over two hours with approximately one-minute cadence shows the longitudinal magnetic field evolution associated with the BP between 22 and 24 hours into its lifetime. High-resolution MDI data are required to sufficiently resolve the small-scale magnetic structure of the BP. Unfortunately, only data from these two hours are available. The XRT observations show that the overall structure of the BP does not change significantly throughout its lifetime, although smaller changes are occurring (see later).

The magnetic field underlying the BP is composed of a region of positive magnetic polarity of about 10 arcsec$^2$ in size, and a larger region of negative polarity in the shape of an "upside-down question mark" (20 arcsec in diameter, see Fig. 1). During the two-hour observation period, the variation of the integrated magnetic flux within both polarities was relatively small, as shown in Fig. 4.

We modelled the three-dimensional magnetic topology (Démoulin & Priest 1992; Longcope & Klapper 2002; Beveridge et al. 2002; Longcope 2005) of the BP and a 150″× 150″ region surrounding it. Ideally the largest area possible would be selected (*i.e.* extending out to where the data starts to become questionable near the limb), so as to include the effects of strong sources far from the feature of interest. However, there is a compromise to be made here, because the topological reconstruction method that we use requires the selected area to be small enough to be accurately represented by a set of point sources on a flat photospheric plane. This means that the selected region must be small enough to make the assumption of a flat photospheric boundary acceptable. The box size was chosen on the basis of both these considerations. Additionally as a check, the topology was also calculated for a 100″× 100″ region centred on the bright point, and although some features shifted position slightly, the large-scale topological structure was identical. We therefore have a high level of confidence that the BP topology that we present here is a good approximation to the true magnetic field configuration that occurred in the solar atmosphere.

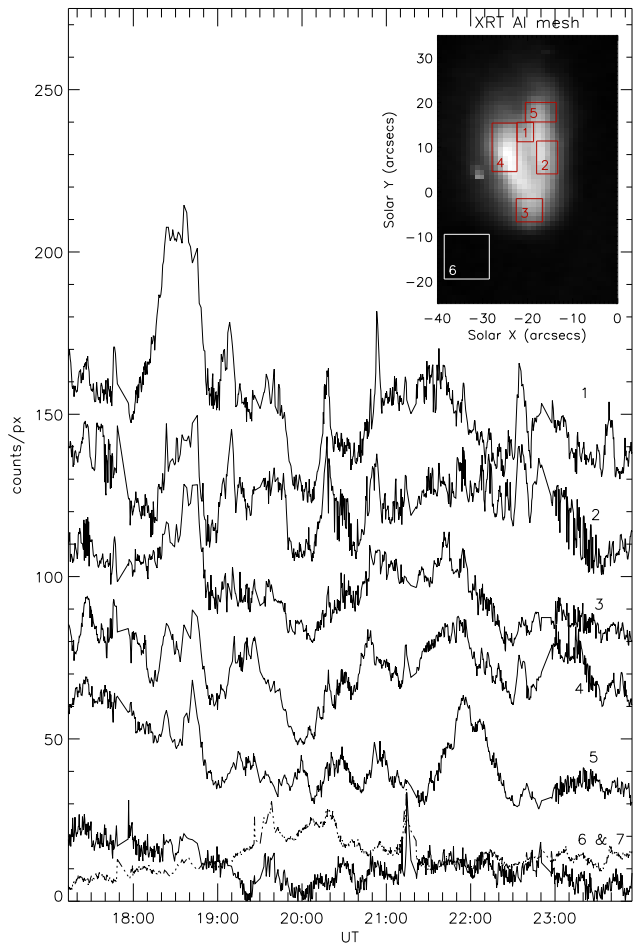

**Fig. 7:** XRT lightcurves at five locations within the BP (1-5) and two outside the BP; the 6$^{th}$ is in the BP neighborhood and the 7$^{th}$ is in a quiet area of the corona (60″ away from the BP). Each lightcurve has been offset in the vertical in order to show the variability; the 6$^{th}$ and 7$^{th}$ curves have been multiplied by a factor of 25 for a clearer view.

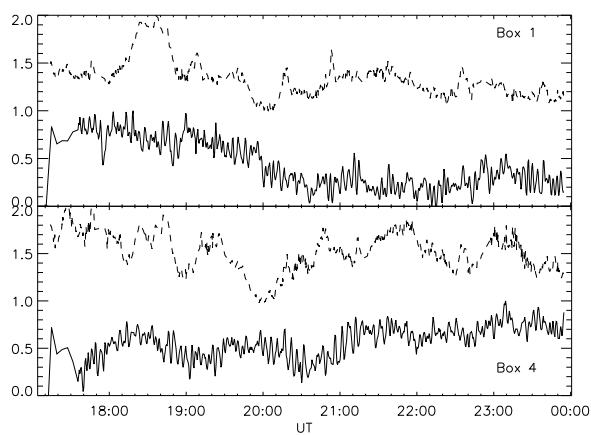

**Fig. 8:** Normalised light curves from Ca II H (solid curve) compared with the corresponding XRT light curves (dashed curve), for two of the boxes in the Figure 7.

First a five-minute running average of the raw magnetograms was calculated, to account for the ubiquitous five-minute oscillations, and to reduce high-frequency noise. We then used the YAFTA feature-tracking code of Welsch et al. (2004) & DeForest et al. (2007) to find and track features in the magnetogram. These features were further reduced to point magnetic sources, with location and strength determined by their parent features. The MPOLE topology code of Longcope (1996) was used to extrapolate the 3D potential magnetic field created by the point sources, and to determine its magnetic topology, *i.e.* the locations of the magnetic null points, spine field-lines, separatrix surfaces, and separators.

Because the positive polarity region (P) is so much weaker than the negative polarity region (N), all of the magnetic flux from the positive polarity connects to the negative polarity. The overlying field is produced by two plage regions, one dominated by negative magnetic polarity to the east and slightly south of the BP, and the other dominated by positive magnetic polarity to the west and slightly north of the BP. This means that field-lines from the negative polarity that do not connect to the positive polarity have a tendency to connect to source regions in the positive plage area. Such field-lines can extend at least 100 arcsec away from the BP, and probably much further. This means that the BP is magnetically linked to many other features far away from it on the solar surface. The large-scale 3D magnetic field structure of the BP is illustrated in Figure 5 – it consists of a large dome of magnetic flux, that contains all the field-lines that are observed to brighten in XRT. The structure is similar to the BP studied by Maclean et al. (2008a). On a large scale, the topology is in an intersecting state (Beveridge et al. 2004). However, in this BP, the brightening is only observed in one of the flux domains (P–N). This is likely to be because these field-lines are much shorter than their counterparts in the other flux domain, so the same amount of heating leads to a stronger brightening.

Figure 6 shows the photospheric footprint of the magnetic topology from Figure 5 at 21:14:02 UT. Thanks to the high-resolution magnetogram data, and as can be seen in Figure 1, it is clear that both the positive and negative polarities are made up of smaller flux concentrations. The positive region consists of 3 small flux concentrations (each modeled here as a point magnetic source), while the negative flux region consists of at least 8 small flux concentrations. The exact number of small flux concentrations detected depends largely on the detection threshold selected to define and track the magnetic features. However, while a range of suitable threshold values exists, in fact whichever value is chosen the number and size of these small flux concentrations within the BP turn out to vary on a much shorter timescale than the apparent overall magnetic structure of the BP. While the BP exists for 37 hours without remarkable changes in its large-scale topological structure, the small-scale structure is varying on a time-scale of minutes.

This short-term variability can also be seen in the light-curves of the X-ray flux. For example, Figure 7 shows five locations within the BP as seen in XRT. Most of these locations show periods of short-term (i.e. several minutes) brightenings. In many instances, these brightenings are similar in the different locations. However, the observed variation in emission on short timescales suggests that the reconnection process is unsteady, although the global picture of emission in the X-rays is very well reproduced by the simple potential field model. Light-curves of the same regions taken in Ca II H do not show any similarity with the XRT data (Figure 8).

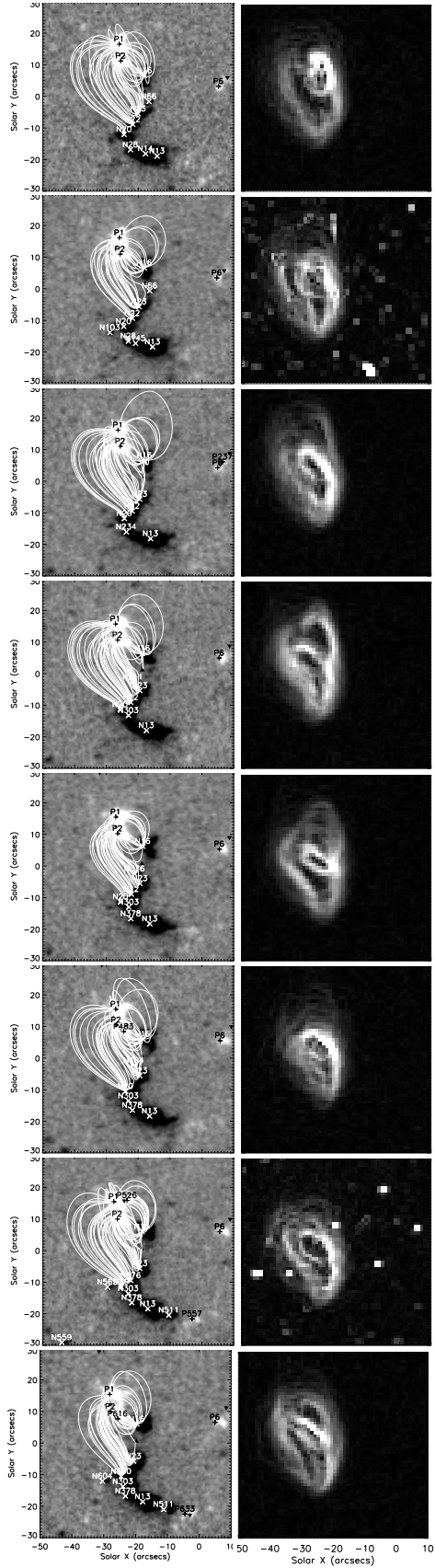

**Fig. 9:** Comparison of MDI possible loop with the image processed by edge detection from XRT, the first one is at 20:54 UT and thereafter every 20 minutes. A movie sequence of this can be viewed online, see movie4.

It is however remarkable that this continual succession of small flux concentrations should emerge and develop in such a way as to maintain the large-scale structure of the BP over several hours to days. Similar behaviour is also observed in the BP studied by Maclean et al. (2008b). Figure 9 shows a comparison between MDI with the field-lines derived via the extrapolation outlined above and XRT images processed by an edge detection routine (Berry & Burnell 2005; Mattis & Kimball 2007); Most of the loops as seen in the XRT X-ray images are in excellent agreement with the field-lines, some small loops are however not observable in the extrapolated data, perhaps implying that they result from a non-potential configuration. Also, there are field-lines outlined in the extrapolation which are not seen in the X-ray images. This is not surprising as these may not be at a temperature which is well observed in the XRT Al-mesh filter. This filter has its peak sensitivity at ≈6MK, with a second less sensitive peak below 1MK. An additional consideration is that the plasma density may be low, hence the loop is not detectable. However, the overall impression from the comparison of the extrapolated field-lines and the X-ray images is that a large fraction of the bright-point at this time is well represented by a potential model.

To our knowledge, this is the first time that the fine loop structure of a BP has been resolved in X-ray data and compared with the extrapolated magnetic field configuration. Assuming that no strong electric currents exist in the quiet Sun, using a potential field extrapolation is a reasonable first step towards resolving the magnetic topology. However, it has recently been shown by Santos et al. (2008) that brightenings in the solar corona associated with BPs are related with the strongest electric current concentrations. Therefore, further work is required using linear force-free extrapolation.

Linear force-free models require a free parameter $\alpha$ which can be specified by a fitting procedure (e.g. from EUV-images) as described in Wiegelmann et al. (2005). The method has been applied so far only to active regions. A forthcoming work will use this method as it has previously done for active regions (Marsch et al. 2004) and compare the obtained magnetic topology of BPs with their EUV and X-ray fine loop structure.

Nonlinear force-free models have been applied only to ARs so far. This method requires vector magnetograph data and, in particular, accurate measurements of the transversal magnetic field component with a sufficient large FOV. In quiet Sun regions the existing measurements of the transversal magnetic field component have a signal-to-noise ratio that is far too low for meaningful nonlinear force-free field extrapolations.

### 4.1. Correlation Analysis

We performed a correlation analysis to find out if there is any relationship between the XRT light curve and the variations in the integrated positive and negative magnetic fluxes from the BP (shown in Figure 4). Flux emergence and cancellation are both known to play a role in coronal heating (Low 1996; Archontis et al. 2005; Priest et al. 1994; von Rekowski et al. 2006b), and so we wished to investigate whether this relationship could be observationally confirmed, with changes in the BP magnetic flux triggering X-ray brightenings.

To create the flux timeseries (Fig. 4), we define a threshold value of ∼ 150 G for the magnetogram data. The value for each frame was then given by the total flux from all pixels with magnetic field strength greater than the threshold value. For the XRT flux we integrated all the flux inside a box of $40'' \times 50''$.

The linear Pearson correlation coefficient describes how well two datasets are correlated. It can vary between −1 and 1, where values close to 1 mean that a strong correlation exists, values close to −1 mean strong anti-correlation, and values close to 0 indicate that there is no correlation between the two datasets. Comparing the XRT light-curve taken over the whole BP and the total integrated positive magnetic flux gives a value of 0.746882, so there is a strong correlation. On the other hand, comparing the total XRT light-curve and the total integrated negative magnetic flux gives a value of −0.0181146; no correlation exists.

Given the correlation of the positive BP magnetic flux with the XRT light-curve (Figure 9), we wished to determine if there was any time lag between peaks in the two datasets, possibly indicating how long it takes for the effects of flux emergence/cancellation to show up as brightenings in X-rays.

Figure 10 shows the results of a cross-correlation analysis. There is no significant correlation for the negative magnetic flux, as expected. For the positive magnetic flux, the best correlation is found to occur with a lag time of zero minutes, *i.e.* instantaneously. This means that energy released by reconnection stimulated by flux emergence or cancellation in the positive magnetic polarity region, shows up within one minute (our cadence time) in X-rays.

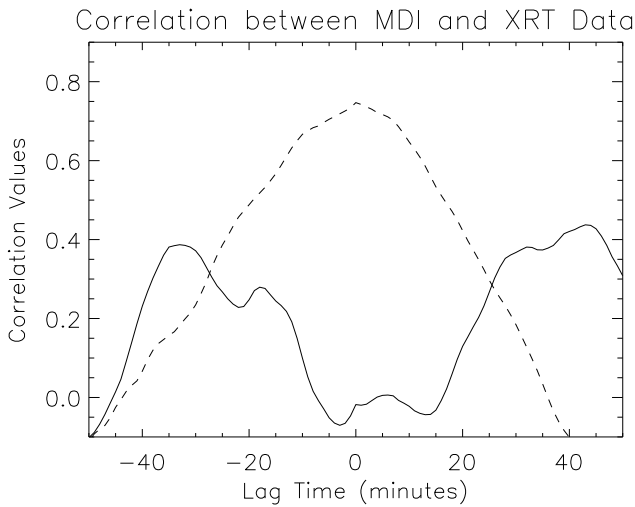

**Fig. 10:** Values of the cross-correlation coefficients between the XRT lightcurve and the positive (dashed line) and negative (solid line) integrated magnetic fluxes from Figure 4, for a range of lag times.

### 4.2. Small Chromospheric Bright Points

Recently Otsuji et al. (2007) with Hinode observations and previously de Wijn et al. (2005) and Sánchez Almeida et al. (2004) with ground-based observations showed that brightenings in Ca II images are related to the concentration of magnetic field as can also be seen by comparison of our MDI and Ca II data. (movie3). We show in Figure 11 the extrapolated field-lines obtained from the MDI data compared to the brightenings in the Ca II images. Each foot-point patch in the MDI data comprises several brightenings when viewed with the higher-resolution Ca II images. The left panel shows how the northern P1 polarity connects to the negative flux patches, while the right panel shows the connections of the southern P2 pole. P1 is connected with five negative patches while P2 is connected with three sources.

The difference between the loop sizes from P1 and P2 is because P2 is closer than P1 to the negative sources. This shows an excellent correspondence between the footpoints of the extrapolated field-lines and the Ca II BPs.

## 5. Discussion & Conclusions

The bright point, as viewed in the X-ray images, showed a structure of $20'' \times 40''$, with several clearly resolvable loops. Its total lifetime was ≈ 37 hours. As seen in higher resolution, the BP's chromospheric structure comprises around 50 small bright point-like features each with a size of $1'' \times 1''$. There are some small points that belong to larger features while others are isolated. These latter ones are more dynamic. When these images are compared with those obtained in the G-Band, some of them are perfectly correlated, but in some cases when a feature is bright in the Ca II H line, it is dark in the G-Band. The Ca II BPs are a good indicator of enhanced magnetic activity. Furthermore, they are the locations of the footpoints of the X-ray loops.

The spectroscopic data showed a Doppler shift of several km s$^{-1}$ in both coronal and transition region lines. This is also the location of the largest line broadening. These values are in the same range as those derived by Brosius et al. (2007) who found, with EUNIS observations, Doppler shifts ranging from −14 to +14 km s$^{-1}$ in Mg IX 368.1 Å, and from −26 to +35 km s$^{-1}$ in the hotter Fe XIV and Fe XVI lines. Excellent agreement was obtained between the two different electron density line ratios, implying an electron density of ≈ $3.7 \times 10^9$ cm$^{-3}$.

The magnetic structure of the BP consists of one region of positive magnetic polarity of ≈10 arcsec$^2$ in size, and about 10 arcsec south of this on the solar disk, another larger region of negative polarity of ≈20 arcsec in diameter. An excellent correlation was found between the positive magnetic flux and the X-ray flux with a lag time of zero minutes. The remarkable agreement between the extrapolated magnetic field configuration and the X-ray images suggest that the overall BP magnetic structure is close to potential, although small time-scale brightenings occur at various locations throughout the bright point.

We postulate that the positive magnetic flux is well-correlated with the total XRT light-curve because the positive source region is so much weaker than the negative source region. The XRT brightenings all occur along relatively short field-lines that are linked to the positive source, so any flux emergence or cancellation in the positive polarity region will have a direct effect on the coronal bright point. Contrast this with the negative polarity region – many of its associated field-lines do not form part of the coronal bright point, and do not even link to the positive polarity region. Any heating caused by flux emergence or cancellation here may be spread thinly along field-lines that connect far from the core of the coronal bright point, so a strong correlation with the total XRT light-curve should not be expected. Moreover there is a very good correlation between the small BPs that appear in the chromosphere with the foot-points of the loops within the coronal BP, demonstrating that those brightenings are produced by a concentration of the magnetic field.

In a recent loop study, Aulanier et al. (2007) showed that slip reconnection should be considered. There is evidence of reconnection here (and in real life it probably is slip reconnection), but unfortunately, given the scale of the loops in the present data and the spatial resolution of the data, we are unable to comment on its importance here. In our reconstruction of the field lines we are looking at real magnetic nulls and separatrix surfaces, rather than quasi-separatrix layers where slip reconnection occurs. Also, the limitations of the modelling technique mean that we are unable to use our potential field extrapolation to try to estimate a realistic reconnection rate.

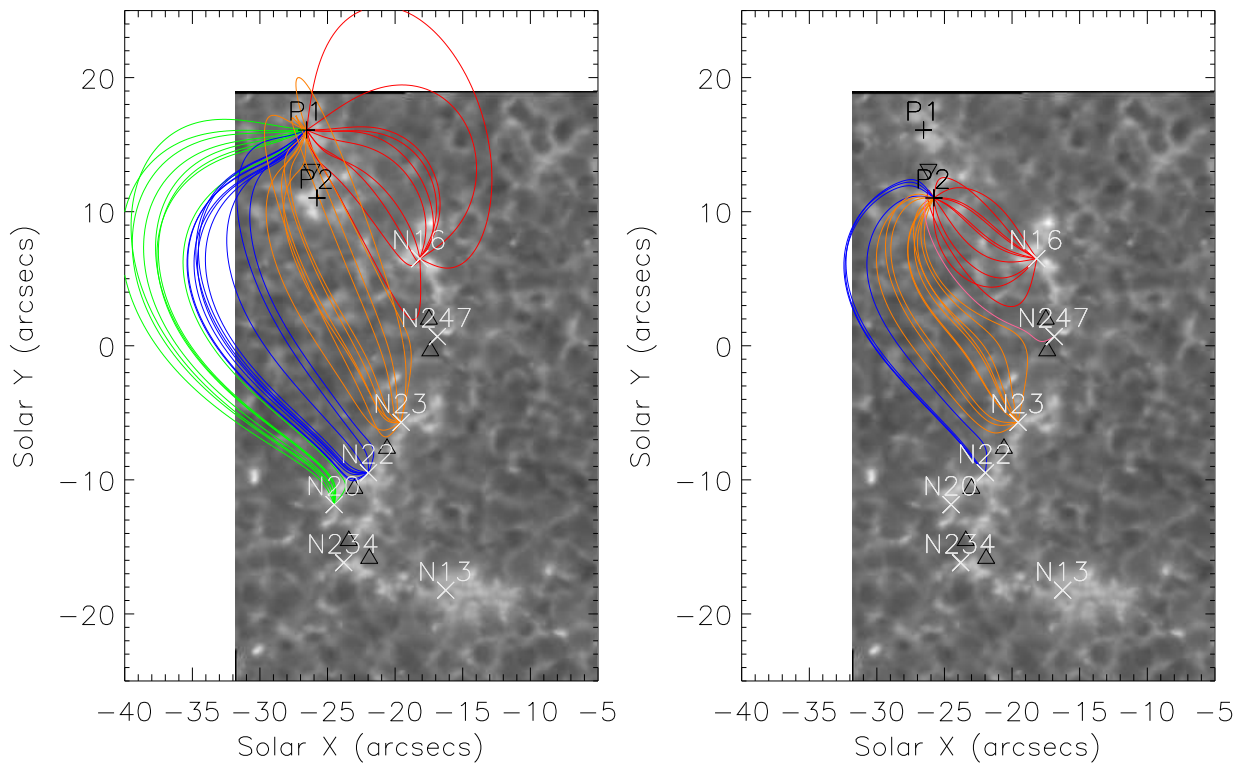

**Fig. 11:** Ca II images at 21:37 UT with the magnetic field lines overplotted. In the left panel, the field-lines originate in the northern positive polarity, with different colours for the different foot-points. In the right panel, the field lines originate in the more southerly positive pole. There is a correspondence between the foot-points and the small bright points (see the third panel in Fig 9 for the corresponding XRT image).

Further work is required to determine whether the results obtained here are applicable to the majority of observed solar coronal bright points, and how a BP's magnetic topological structure evolves throughout its lifetime.

*Acknowledgements.* We would like to thank the Hinode SOT, EIS & XRT teams plus the SoHO SUMER, MDI & EIT teams for their help in obtaining and reducing the data. SUMER, EIT and MDI are part of SOHO, the Solar and Heliospheric Observatory, which is a project of international cooperation between ESA and NASA. Hinode is a Japanese mission developed and launched by ISAS/JAXA, collaborating with NAOJ as a domestic partner, NASA and STFC (UK) as international partners. Support for the post-launch operation is provided by JAXA and NAOJ (Japan), STFC (UK), NASA, ESA, and NSC (Norway). CHIANTI is a collaborative project involving the NRL (USA), RAL (UK), MSSL (UK), the Universities of Florence (Italy) and Cambridge (UK), and George Mason University (USA). We would like to thank to the referee, Scott McIntosh, for his very valuable comments and insight on an earlier draft and Thomas Wiegelmann for reading and commenting on the paper.

## Movies

**Movie1:** http://star.arm.ac.uk/preprints/2008/533/fig1.html
**Movie2:** http://star.arm.ac.uk/preprints/2008/533/fig2.html
**Movie3:** http://star.arm.ac.uk/preprints/2008/533/fig2b.html
**Movie4:** http://star.arm.ac.uk/preprints/2008/533/fig10.html

3